\documentclass[prl,aps,twocolumn,floats,showpacspsfig]{revtex4} 
\usepackage{epsfig,amsfonts}
\def\fd{{\phantom{\dagger}}}
\def\bbI{{\mathbb I}}

\def\be{\begin{equation}}
\def\ee{\end{equation}}
\def\bdm{\begin{displaymath}}
\def\edm{\end{displaymath}}
\def\bea{\begin{eqnarray}}
\def\eea{\end{eqnarray}}

\newcommand{\p}{\partial}

\newcommand{\rd}{\mbox{d}}
\newcommand{\ri}{\mbox{i}}
\newcommand{\re}{\mbox{e}}

\begin{document}
\title{Analytical results for the 
Coqblin-Schrieffer model\\
 with generalized magnetic fields }
\author{V.V. Bazhanov$^{1,2}$, S.L. Lukyanov$^{3,4}$ and A.M. Tsvelik$^5$}
\affiliation{$^1$ Department of Theoretical Physics, 
Research School of Physical Sciences and Engineering, \\
Australian National University, Canberra, ACT 0200, Australia\\
$^{2}$ Research Institute for Mathematical Sciences, Kyoto 
University, Kyoto 606-8502, Japan.\\
$^{3}$ Department of  Physics and Astronomy, 
Rutgers University, Piscataway, NJ 08855-0849, USA,\\
$^4$L.D. Landau Institute for Theoretical Physics, Chernogolovka,
142432, Russia,\\ 
$^5$ Department of  Physics, Brookhaven 
National Laboratory, Upton, NY 11973-5000, USA}
\date{May 8, 2003}

\bigskip

\begin{abstract}
Using the  approach alternative to the traditional 
Thermodynamic Bethe Ansatz, we derive 
analytical expressions for the free energy of Coqblin-Schrieffer 
model with  arbitrary magnetic and  crystal fields. 
In  Appendix we discuss two concrete 
examples including the  field  
generated crossover from the $SU(4)$ to  
the $SU(2)$ symmetry in the $SU(4)$-symmetric model.

\end{abstract}
\maketitle

The advances of  nanotechnology 
have given an additional weight to the 
old problem of Kondo effect. 
Artificially manufactured structures such as 
quantum dots emulate the behavior of ``natural'' magnetic 
impurities though on  different energy scales. 
Using  technological means one can 
widely vary parameters of the dots 
thus getting  an  access to 
previously experimentally 
unexplored regions of the phase diagram. 

The most physically transparent 
situation corresponds to 
the case when magnetic impurity 
(or quantum dot) has a perfect symmetry. 
This, however, is rarely achieved 
in real systems due to a presence of  
the crystalline lattice. 
Let us consider, for instance, 
magnetic impurities made of   rare-earth magnetic
ions of Ce and Yb. In the presence of a strong
spin-orbital coupling an  $f^1$(Ce) or $f^{13}$(Yb)-orbital
is characterized by the total angular
momentum $j$ ($j =5/2$ for Ce and 7/2 for Yb) such
that  an isolated ion has the
$SU(n)$ symmetry with  $n = 2j + 1$.
In the crystalline environment  this symmetry
is broken. The interplay of 
these effects with the Kondo screening can be studied using   
the Coqblin-Schrieffer model\ \cite{coqblin}:
\bea
\label{ksiuy}
{\bf H} = \sum_{k,a}
k {\bf c}^\dagger_{k,a}{\bf c}^\fd_{k,a} + 
\frac{J}{V}\sum_{k,k'\atop a,b}{\bf c}^\dagger_{k,a}
{\bf c}^\fd_{k',b} {\bf f}^\dagger_{b} {\bf f}^\fd_a
 +\sum_{a}
H_a {\bf f}_a^\dagger {\bf f}^\fd_a
\eea
where ${\bf c}^\dagger_{k,a} ,\, {\bf c}^\fd_{k, a} $
are creation (annihilation)
operators of the conduction 
electrons partial
harmonics with the angular
momentum projection
$m=j+1- a\ (a=1,\ldots n)$;
\  ${\bf f}^\dagger_a$ and ${\bf f}_a$-operators describe
the impurity spin and $V$ is the volume of the system. The generalized
magnetic field $H_a$ originates from crystal fields inherent to the
material and the external magnetic field.  
Since Hamiltonian (\ref{ksiuy}) commutes with the operator
$${\bf q}=\sum_{a=1}^n\, {\bf f}_a^\dagger {\bf f}^\fd_a \ ,$$
  one can  assume without loss of generality that
\bea
\label{snsy}
\sum_{a=1}^nH_a=0\ .
\eea
Notice that the  cases described
above correspond to the sector  of the Coqblin-Schrieffer model
with the occupation number $q=1$.

Since model (\ref{ksiuy}) has a linear spectrum, 
it  has to be equipped 
with the ultraviolet (UV) cutoff $\Lambda$ and
a consistent removal of the UV divergences requires that
the ``bare'' coupling $g_0=n J \rho(0)$
(here $\rho(0)$ is the conduction
electron density of states
at the chemical potential)
be given certain dependence of the
cutoff momenta (see, e.g. \cite{read}):
\be
\label{ksmnddh}
\Lambda\ {\rd g_0\over \rd\Lambda}=-g_0^2+{g_0^3\over n}+\ldots\ .
\ee
Eq.(\ref{ksmnddh}) shows that for
positive $g_0$
the Coqblin-Schrieffer  model   acquires a physical
energy scale, the Kondo temperature 
\be
\label{ksjdya}
T_K\sim \Lambda\ g_0^{1\over n}\re^{-{1\over g_0}}\ ,
\ee
and renormalization trades the bare  coupling
constant $g_0$ for the renormalization group 
invariant scale $T_K$. Therefore the
partition functions of the model in the sector
with a given  occupation number $q$,  $Z_q$,  actually depends
on the dimensionless  combinations
$T/T_K$\ and\ $H_a/T_K$.
Of course,  formula (\ref{ksjdya}) does not
specify the physical  energy scale
uniquely and  in order  to define
$T_K$ unambiguously
we shall impose the  conventional  normalization condition \cite{wtsv},
\be
\label{lsdjku}
\lim_{T\to 0}\, C_{1}(T)/T\big|_{H_a=0}=
 {\pi\over 3}\ {n-1\over T_K}\ ,
\ee
where $C_{1}(T)=T{\partial^2\over \partial T^2} (T\log Z_{q=1})$ 
is the heat capacity
of the impurity in the sector  with the occupation number $q=1$.

If the fields are weak in comparison with 
the Kondo temperature (\ref{lsdjku}),
the behavior of the system is
governed by $T_K^{(n)}$ (the Kondo temperature of the fully 
$SU(n)$-symmetric model) which for 
rare earth impurities may be as large as
several hundred degrees. 
On the other extreme, 
if the generalized magnetic  fields 
exceed $T_K^{(n)}$ and break down the 
degeneracy to one Kramers doublet, 
one gets the Kondo temperature of the order of several degrees. 
As an illustration  let us estimate the new 
Kondo temperature for the case when the original $SU(n)$ symmetry 
is broken down to $SU(m)$ by the  fields $H_1, ...H_{n-m} \gg T_K^{(n)}$. 
The estimate is easy if  all fields $H_1,...H_{n-m}$ are of the 
same order ${\bar H}$. Since $\ln T_K^{(n)} \sim -1/nJ\rho(0), ~~ 
\ln T_K^{(m)} \sim -1/mJ\rho(0)$ 
and the dimensionless ground state 
energy depends only on $H_a/T_K^{(n)}$, we 
conclude that the resulting Kondo temperature is 
\be
\label{lsaaadjku}
T_K^{(m)} \sim  {\bar H}\, (T_K^{(n)}/{\bar H})^{n\over m}\ ,
\ee
which may easily constitute a scale vastly 
different from the Kondo temperature of the unperturbed model. 
It is clear that the detailed behavior in 
this crossover interval depends on the field ratios $H_a/H_b$ and 
it would be highly desirable to have analytic tools to 
handle a situation with an arbitrary pattern of fields. 

Historically the thermodynamics of the Coqblin-Schrieffer model has
been examined by the method of Thermodynamic Bethe Ansatz (TBA)
\cite{wtsv,aflo,rasul,coleman,pedroo}.  Unfortunately the TBA
equations corresponding to Eq.(\ref{ksiuy}) are rather complicated to
be studied analytically, and until now the majority of results have
been obtained by means of their numerical integration (see\
\cite{pedro} and references therein).

This work is based on the approach alternative to the TBA. Here we
just outline major steps of our analysis and refer the reader to the
papers\ \cite{blz,bazh,dorey,zam}\ where the method itself was developed.
Let us make some remarks on the
general spirit of the approach.  It uses the fact that integrable
impurity models in general and the Coqblin-Schrieffer model in
particular can be mapped to the (1+1)-dimensional bulk conformal field
theory (CFT) with a non-conformal boundary interaction (see e.g.\
\cite{affleck}).  From this point of view, Eq.(\ref{ksiuy}) belongs to
the class of exactly solvable boundary theories such that the associated
boundary state commutes with the infinite set of mutually commutative
local integrals of motion of the bulk system.  As a matter of fact,
one may say that the boundary state ``generates'' this set in the sense
that it admits the asymptotic large-distance expansion in terms of
these local integrals.  More precisely, the Hilbert space of the bulk
CFT associated with Eq.(\ref{ksiuy}) can be classified in accordance
with the $WA_{n-1}$-algebra \cite{fat} with the central charge $c=n-1$ and
the corresponding boundary state commutes with the set of local
integrals of motion introduced in \cite{fat}.

In the   approach adopted in this paper 
the amplitudes of the
boundary state are related to monodromy characteristics of certain
ordinary
linear differential equations. 
The key ingredient is the equation  of the form,
\bea
\label{smsdhdty}
\Big\{\,  \big(-\ri\, \partial_v+ h_1)\ldots
(-\ri\, \partial_v+ h_n)-\re^{n\theta}\ \re^{v} \, v
\, \Big\}\, \Psi=0\  ,
\eea
where  $\theta$ and $h_a$ are  some
(complex) parameters. 
For $\Im m (\theta)=\pi/2$ 
the equation
admits a   solution  which 
specified unambiguously by the following 
asymptotic as $v\to+\infty$:
\bea
\label{smsudu}
\Psi_0(v,\theta) 
\to &&
 \big(-\ri\, 
\re^{\theta}\, v^{1\over n}
\re^{v\over n}\big)^{-{n-1\over 2}}\nonumber \\ &&
 \times\exp\bigg\{\ri\, \re^{\theta}\, \Big({\cal C}
+ \int_{0}^vdu\, u^{1\over n}\,
\re^{{u\over n}}
\, \Big)\, \bigg\}\ . 
\eea
Here  ${\cal C}$ is an arbitrary constant, 
whose  explicit value is not essential for our purposes.
Notice that Eq.(\ref{smsdhdty})
is invariant under the transformation,
$\theta\to\theta+2\pi \ri/ n$, whereas  
the  asymptotic (\ref{smsudu}) is not. Hence
the analytic continuations of $\Psi_0$,
\be
\label{skdny}
\Psi_q(v,\theta)=\Psi_0(v,\theta+2\pi \ri\, q/ n)\ ,
\ee
with  integers $q$,
generate new solutions of Eq.(\ref{smsdhdty}). It is possible to 
show  that the Wronskian 
$W[\Psi_0,\, \Psi_1,\,\ldots \Psi_{n-1}]$
does  not vanish, 
so that the set\ 
$\{\Psi_q\}_{q=0}^{n-1}$\ is a 
fundamental system of solutions of Eq.(\ref{smsdhdty}). By virtue of 
this fact, solution (\ref{skdny}) with $q=n$ can be decomposed as
\bea
\label{sdngu}
\Psi_{n}(v,\theta)=\sum_{q=0}^{n-1} (-1)^{n-q-1}\ Z_q(\theta
+{\ri\pi q/ n})\
\Psi_{q}(v,\theta)\, .
\eea
Following the line of arguments similar to that of\ \cite{blz,bazh},
it is possible to show that if the parameters 
$\theta$ and $h_a$ are identified with
the dimensionless parameters of the Coqblin-Schrieffer model,
\be
\label{sksduyu}\re^{\theta}={1\over 2 n^{1\over n} \Gamma(
1/n)}\ {T_{K}\over T}\, ,\ \ \ \ \ \ h_a={H_a\over 2\pi T}\ ,
\ee
then the function $Z_q(\theta)$ appearing in (\ref{sdngu}) 
coincides with the analytic continuation
of the partition function  of this model, $Z_q$, 
for the sector with the occupation number  $q$. 

The subject of our current 
interest is  the  free energies
\be
\label{smsdjduy}{\cal F}_q=-T\, \log(Z_q)\ 
\ee
at the low temperature limit.
In particular we study
vacuum energies ${\cal E}_q=
{\cal F}_q|_{T=0}$.
They can be obtained from  the semi-classical (WKB) approximation for
Eq.(\ref{smsdhdty}).
The leading
terms in the WKB expansion of the solution $\Psi_0$ (\ref{smsudu})
read,
\bea
\label{msjuy}
&&\Psi_0(v,\theta)\simeq  \Big(-\ri\, X\big(
\re^{\theta}\, v^{1\over n}\ \re^{v\over n}\big)
\Big)^{-{n-1\over 2}}\times\nonumber
\\ &&{}\  \exp\bigg\{\ri\, \re^{\theta}\, \Big({\cal C}+
\int_{0}^v \rd u\, u^{1\over n}\, 
\re^{{u\over n}}\, \Big)
-\nonumber \\ &&{}\  
\ri\, \int^{+\infty}_v\rd u\, \big(X(Y)-
Y\big)\big|_
{Y=\re^{\theta}\, u^{1\over n}\ \re^{u\over n}}\, \bigg\}\, .
\eea
Here  $X=X(Y)$ is the  solution of
the algebraic equation
\be
\label{hsdyyt}Y^n=(X+h_1)
\ldots (X+h_n)\ , 
\ee
such that 
\be
\label{saakiuy}X(Y)\to Y\ \ {\rm  as}\ \ \ \ 
Y\to\infty\ .
\ee
In fact, $X=X(Y)$ is a multi-valued function of 
the  argument $Y$ and Eq.(\ref{saakiuy}) uniquely 
specifies its branch for a sufficiently large $Y$.
The latter solution   admits a convergent  $1/Y$-power
series expansion found by Lagrange\ \cite{lagrange}:
\be
\label{smdyy}X(Y)=Y+{1\over n}\ \sum_{k=1\atop
k\not= 0 (mod\ n)
}^{\infty}
\ I_{k}(h_1,\ldots h_n)\ Y^{-k}
\ .
\ee
Here $I_k$ are symmetric polynomials given by
\bea
\label{msndg}
&&I_k
(h_1,\ldots h_n)
= 
\sum_{\alpha_1,\ldots\alpha_{n-1}\geq 0\atop
2\alpha_{1}+3\alpha_{2}+\ldots n\alpha_{n-1}=k+1}
\frac{(-1)^{\alpha_1+
\ldots +\alpha_{n-1}}}
{\alpha_1!\alpha_2!\ldots \alpha_{n-1}!} \nonumber\\
&&\times\frac{\Gamma(\alpha_1+
\ldots+\alpha_{n-1}-{k\over n})}
{\Gamma(1-{k\over n})}
\ G_2^{\alpha_{1}}\, G_3^{\alpha_{2}}\,
\ldots G_n^{\alpha_{n-1}}\, ,
\eea
where we  use the following notation  for the
elementary  symmetric polynomials $G_k$:
\bea
\label{lskuu}G_{k}
=\sum_{1\leq a_1<a_2\ldots <a_k\leq n}h_{a_1}\ldots h_{a_k}
\ .
\eea
Note that according to the constraint (\ref{snsy}),
we have set $G_1=0$, and also
$G_2=-{1\over 2}\ \sum_{a=1}^n h_a^2.$

Using the Lagrange formula\ (\ref{smdyy}) and (\ref{msjuy})
it is possible to show that the
vacuum  energies  admit the following
weak-field expansion\ \cite{Mellin}
\bea
\label{jsdht}
&&{\cal E}_q=E_0 \sin\Big({\pi q\over n}\Big)+
{T_K\over n} \times\\
&&\sum_{k=1\atop
k\not= 0 (mod\ n)
}^\infty\, C_k\,
\sin\Big({\pi k q\over n}\Big)\
I_k\left( {H_1\over 2\pi T_K},\ldots 
{H_n\over 2\pi T_K} \right)\, ,\nonumber 
\eea
where
$$C_k=2^{k+1}\ k^{k/ n}\
\Gamma^k(1/ n)\,  \Gamma(-k/ n)\ .$$
The first term in (\ref{jsdht}) is the vacuum
energy for  zero fields and the value of $E_0\sim T_K$
is related to  the  choice 
of the constant ${\cal C}$ in (\ref{smsudu})\ \footnote{
According to  the $1/n$-expansion of Ref.\cite{read},
$E_0/T_K=1+O(n^{-2}).$}. 
Notice that $I_k$ (\ref{msndg}) are 
homogeneous symmetric  polynomials of the variables
$h_a$ of the degree $k+1$, i.e.
\bea
\label{juyt}I_k(\lambda h_1,\ldots \lambda h_n)=\lambda^{k+1}\ 
I_k(h_1,\ldots h_n)\ ,
\eea
so Eq.(\ref{jsdht}) can be  considered as a power series expansion
in ${\bar H}/(2\pi T_K)$, where 
${\bar H}=\sqrt{{1\over n}\sum_{a=1}^nH_a^2}$.
Thus if the fields are weak in comparison with the 
Kondo temperature 
the behavior of the system is
governed by the convergent series (\ref{jsdht}).
At the same time this series has a finite convergence radius
and  defines the  multi-valued function
of complex variable $\bar H$\ (see  Appendix for  examples).

We have also examined the structure 
of the low temperature expansion
of the free energy (\ref{smsdjduy}) and found the 
following asymptotic series:
\bea
\label{jaasdht}
&&{\cal F}_q=E_0\sin\Big({\pi q\over n}\Big)
+{T_K\over n}\
\sum_{k=1\atop
k\not= 0 (mod\ n)
}^\infty\, C_k\,
\sin\Big({\pi k q\over n}\Big)\nonumber\\
&&{}\ \times\bbI_k\left( {H_1\over 2\pi T},\ldots
{H_n\over 2\pi T} \right) \, \left({T\over T_K}\right)^{k+1}
+O(T^{\infty}) \ .
\eea
Here the numerical coefficients $C_k$ are the same as
in (\ref{jsdht}) and
$\bbI_k(h_1,\ldots,h_n)$ is a symmetric polynomial
of the degree $k+1$  of the form,
\be
\label{slkdi}\bbI_k=I_{k}+J_{k}\ ,
\ee
where
$I_{k}$ is given by Eq.(\ref{msndg}) and 
$J_{k}=J_{k}(h_1,\ldots,h_n)$ 
are  some symmetric polynomials
of the  degree $k$. 
Currently the polynomial $\bbI_k$  are not known in a closed form for arbitrary
$k$. Nevertheless they admit a simple 
algebraic description.  
As has been  mentioned before, the boundary state 
associated with the
Coqblin-Schrieffer model commutes with the infinite set of
mutually commutative integrals of motion  
for the $WA_{n-1}$-algebra  with the central charge
$c=n-1$.
It turns out that the polynomials $\bbI_k$ appearing  in 
Eq.(\ref{jaasdht}) coincide with vacuum
eigenvalues of  these conserved charges.
Therefore they
can be calculated purely algebraically for any finite $k$.
In particular we found,
\bea
\label{skdiuu}
&&\bbI_1=- G_2+{n-1\over 24}\, ,\nonumber\\ &&
\bbI_2=- G_3\, ,\nonumber\\
&&\bbI_3= -G_4+{n-3\over 2 n}\, G_2^2-{n-3\over 8n}\, G_2+
{3(n-1)(n-3)\over 640 n}\, ,\nonumber \\
&&\bbI_4=-G_5+
 {n-4\over  n}\ G_3G_2  -{n-4\over 3n}\ G_3\, .
\eea
Here $G_k$ are the elementary symmetric polynomials (\ref{lskuu}).
Notice that for the physically
interesting  case of the 
magnetic field configuration
\be
\label{ddooh}h_a=h\ (j+1-a)\, ,\ \ \ \ j={n-1\over 2}\ ,
\ee
all polynomials $\bbI_{k}$ with even $k$  vanish,
whereas using Eq.(\ref{skdiuu}) one has:
\bea
\label{skdiioo}
&&\bbI_1={n-1\over 24}\ \big(\, n(n+1)\, h^2+1\, \big)\, ,\nonumber\\
&& \bbI_3={(n-1)(n-3)\over 1920 n}\times\\
&&\ \big(\, 
n^2(n+1)(n+3)\, h^4+10\, n(n+1)\, h^2+9\, \big)\ .\nonumber
\eea

In conclusion let us just repeat 
the basic results of the paper. 
We have outlined a method which allows 
to extract analytical results for the thermodynamics 
of Coqblin-Schrieffer model in the presence 
of magnetic and crystal fields. 
The general formulae drastically simplify for the $SU(4)$ model (see
Appendix).  

\bigskip

{\bf Acknowledgments}. 
S.L.L. and A.M.T. are  grateful to N. Andrei, F.H.L. Essler and
A.B. Zamolodchikov 
for interesting discussions and constructive criticism. 
A.M.T. acknowledges the support from
US DOE under contract
 number DE-AC02-98 CH 10886.
Research of S.L.L. is supported
in part by DOE grant $\#$DE-FG02-96 ER 40959; he also  acknowledges a
support from
Institute for Strongly Correlated and Complex Systems at BNL.

\bigskip
\bigskip
{\bf Appendix}

\bigskip

In this appendix we  illustrate 
the expansion (\ref{jsdht}) using two examples.

Our first example  is related to the case $n=4$.
Let us  consider the following
configuration of
the generalized magnetic  fields:
\bea
\label{sndht}
H_1=-H_4\, ,~~\  H_2=-H_3\, , 
\eea
with
$H_1\geq H_2\geq 0$.
In this case Eq.(\ref{hsdyyt}) is especially simple and
can be written in the form
\be
\label{mskduy}
Y^4=(X^2-h_1^2)(X^2-h_2^2)\, .
\ee
Its solution $X=X(Y)$ satisfying (\ref{saakiuy}) 
reads explicitly:
\be
\label{diuue}
x=y\ \sqrt{\sqrt{1+\epsilon/y^4}+1/ y^2}\ ,
\ee
where $y=Y/{\bar h},\ x=X/{\bar h}$, and
$${\bar h}=\sqrt{{h_1^2+h_2^2\over 2}}\ ,\ \ \ \ 
\epsilon=\Big({h_1^2-h_2^2
\over  h_1^2+h_2^2}\Big)^2\ .$$
All even  coefficients $I_{2l}$ in
 series (\ref{smdyy})  vanish now, whereas $I_{2l-1}$ can
be expressed in terms of the hypergeometric function:
\be
\label{sdkiiu}
I_{2l-1}=2\, (-1)^{l-1}\ {\Gamma(l-{1\over 2})\over
\sqrt{\pi}\, l!}\ {\bar h}^{2l}\ {}_2F_1\Big(-{l\over 2}, {1\over 2}
-{l\over 2},
{3\over 4}-{ l\over 2};\, \epsilon\, \Big). 
\ee
Combining this equation with (\ref{jsdht}) one obtains,
\bea
\label{msndyt}
&&{\cal E}_q=E_0\sin\Big({\pi q\over 4}\Big)
+{T_K\over \sqrt{2}\Gamma(1/4)}\times\\
&& \sum_{l=1}^{\infty}
{(-1)^{l}\over l!}\ (2l-1)^{{l\over 2}-{5\over 4}}\ \Gamma\Big(
{l\over 2}+{1\over 4}\Big)
\ {\sin({\pi (2l-1)q\over 4})\over
\sin({\pi (2l-1)
\over 4})}\times\nonumber\\
&& {}_2F_1\Big(-{l\over 2}, {1\over 2}
-{l\over 2},
{3\over 4}-{ l\over 2};\, \epsilon\, \Big)\ 
\Big(\,{\Gamma({1/ 4})\sqrt{2}
{\bar H}\over \pi T_K}\, \Big)^{2 l}\, ,\nonumber 
\eea
with
$${\bar H}=\sqrt{H_1^2+H_2^2\over 2}\ ,
\ \ \ \  \epsilon=\Big({H_1^2-H_2^2\over H_1^2+H_2^2}\Big)^2
\leq 1\ .$$
Eq.(\ref{msndyt})\ can be rewritten in the
form of a convergent 
integral which is useful for an analytical continuation of the  
power series 
expansion outside 
its convergence disk,
$$|{\bar H}/ T_K|< {\pi \re^{-{1\over 4}}\over \sqrt{2}\Gamma(1/4)}\ 
{\rm min}\Big[\, \epsilon^{-{1\over 4}},\, (1-\epsilon)^{-{1\over 4}}
\Big]\ .$$
Explicitly one has ($q=1,2,3$):
\bea
\label{samdjytr}
&&
{\cal E}_q=E_0\, \sin\Big({\pi q\over 4}\Big)+{\bar H}\ 
\sin\Big({\pi q\over 4}\Big)\times\\
&&\int_{-\infty}^{\infty}{\rd\omega\over 2\pi}\
(\ri\omega+0)^{\ri\omega-1}\ {\Gamma(-{1\over 2}-2 \ri\omega)
\over \Gamma({1\over 2}- \ri\omega)}
\Big({2\Gamma(1/4) {\bar H}\over \pi T_K}\Big)^{4\ri\omega}\times\nonumber\\ 
&&{}_2F_1\Big({1\over 4}-
\ri\omega, -{1\over 4}-\ri\omega,
{1\over 2}-\ri\omega;\,{1\over 2}-(-1)^q\ {1-2\epsilon\over
2}  \, \Big)\, .\nonumber 
\eea
It is particularly illuminating to extract from this expression 
the Kondo temperature for the $SU(2)$ Coqblin-Schrieffer model to 
compare it with Eq.(\ref{lsaaadjku}). The Kondo temperature $T_K^{(2)}$ 
can be  extracted from the magnetic susceptibility 
in the limit when one of the fields is very large. 
Since the energy depends on two fields $H_1,\, H_2$, 
one has to be careful in choosing the right direction of differentiation. 
The right choice of variables is $H_{\pm} = H_1 \pm H_2$ such that 
$\p{\cal E}_1/\p H_{-} =0$ at $H_{-} =0$. 
Then at $H_+\to 2{\bar H}\gg T_K,\  H_- \rightarrow 0$
 we obtain for $q =1$ the following expression for the magnetic susceptibility
\bea
&&\chi =- \frac{\p^2{\cal E}_1}{\p H_-^2}\Big|_{H_- =0\atop H_+=2{\bar H}} \to
{1\over 2\pi\, T_K^{(2)}}\, , \nonumber\\
&&T_K^{(2)} = \frac{\pi^{3/2}}{\sqrt{8} \Gamma^2(1/4)}\
{ T_K^2\over \bar H}\ .
\eea
The above expression for the effective Kondo 
temperature is a particular case of Eq.(\ref{lsaaadjku}) for $n =4,m=2$.

\bigskip

Our second example  is related to the case of an arbitrary integer $n$.
In \cite{wtsv}, the vacuum energy
${\cal E}_1$ was found  for the following field configuration,
\bea
\label{jshy}H_a=\sqrt{2}\ {\bar H}\ 
\cos\Big({\pi(2 a-1)\over 2n }\Big)\, .
\eea
For this pattern the
equation (\ref{hsdyyt})  is  expressed  in terms of
the Chebyshev polynomials $T_n(x)=\cos(n\arccos(x))$:
\be
\label{smdhgt}y^n=2^{1-{n\over 2}}\  T_n(x/\sqrt{2})\, ,
\ee
where $y=Y/{\bar h}$,  $x=X/{\bar h}$.
Therefore
\bea
\label{dhdtrtr}
x=&&
y\ {}_2F_{1}\Big(-{1\over 2n},{n-1\over 2 n},{n-1\over n};\,
{2^{2-n}\over y^{2 n}}\Big)+\nonumber\\
&&{1\over 2y}\ {}_2F_{1}\Big({1\over 2n},
{n+1\over 2 n},{n+1\over n};\,
{2^{2-n}\over y^{2 n}}\Big) \, , 
\eea
and
non-vanishing coefficients in (\ref{smdyy})  and (\ref{jsdht})
are $I_{2 n l-1}\ (l=1,2
\ldots) $ and $I_{2 n l +1}\,
(l=0,1,2,\ldots)$:
\be
\label{skdjdhty}I_{2nl+\sigma}=
\sigma\, {\Gamma({\sigma\over n}+2l)\over l!
\Gamma(1+{\sigma\over n}+l)}\, \Big({{\bar h}
\over \sqrt{2}}\Big)^{2ln+1+\sigma}\, ,
 \  \sigma=\pm 1\, . 
\ee
As in the first example, 
the vacuum   energies (\ref{jsdht}) can be
written in the form of convergent integral \cite{wtsv}:
\bea
\label{dgrew}
&&{\cal E}_q=E_0\sin\Big({\pi q\over n}\Big)+\nonumber\\ &&
\ \ {\sqrt{2}{\bar H}\sin(\pi q/ n)\over 4\pi n}\
\int_{-\infty}^{\infty}{\rd\omega\over 2\pi  }\
(\ri\omega+0)^{\ri\omega-1}\times \\ 
&&\Gamma\Big(-{1\over 2n}-{\ri\omega\over 2} \Big)\, \Gamma
\Big({1\over 2n}-{\ri\omega\over 2}\Big)\ \Big(
{n^{1\over n} \Gamma(1/n){\bar H}
\over \pi \sqrt{2} T_K}\Big)^{\ri n\omega}\, . \nonumber
\eea
It is easy to check that this equation for $n=4$ is 
in agreement with (\ref{samdjytr}) provided $\epsilon={1/ 2}$.

\end{document}